**Tunable, Low Optical Loss Strontium Molybdate Thin Films for Plasmonic Applications**

*Matthew P. Wells*, Bin Zou, Brock Doiron, Rebecca Kilmurray, Andrei P. Mihai, Rupert F. M. Oulton, Lesley F. Cohen, Stefan A. Maier, Neil McN. Alford, Peter K. Petrov*

M. P. Wells, Dr B. Zou, R. Kilmurray, Dr A. P. Mihai, Prof. N. Alford, Dr P. K. Petrov
Imperial College London, Department of Materials, Prince Consort Road, London SW7 2BP, UK
E-mail: m.wells15@imperial.ac.uk
B. Doiron, Dr R. F. M. Oulton, Prof. L. F. Cohen, Prof. S. A. Maier
Imperial College London, Department of Physics, Prince Consort Road, London SW7 2AZ, UK

Strontium molybdate ($SrMoO_3$) thin films are shown to exhibit plasmonic behaviour with a zero crossover wavelength of the real part of the dielectric permittivity tunable between 600 and 950 nm (2.05 eV and 1.31 eV). The films are grown epitaxially on strontium titanate ($SrTiO_3$), magnesium oxide (MgO), and lanthanum aluminate ($LaAlO_3$) substrates by pulsed laser deposition. $SrMoO_3$ is shown to have optical losses lower than those of gold at the point at which the real part of the dielectric permittivity is equal to -2, while possessing low electrical resistivity of 100 $\mu\Omega$ cm at room temperature. Spectroscopic ellipsometry measurements reveal that $SrMoO_3$ shows plasmonic behaviour, moreover we demonstrate that the epsilon near zero (ENZ) wavelength is tunable by engineering the residual strain in the films. The relatively broadband ENZ behaviour observed in $SrMoO_3$ demonstrates its potential suitability for transformation optics along with plasmonic applications in the visible to near infrared spectral range.

**1. Introduction**

Plasmonic phenomena are expected to have a significant impact in applications such as nanoscale imaging,[1] biological sensing,[2,3] energy harvesting,[4,5] and communication systems.[6] As a result of strong coupling between light and electrons at the material surface, plasmonic materials offer a means of overcoming the diffraction limit, thereby enabling manipulation of light at the nanoscale. This allows the high bandwidth capabilities of photonics to be merged with the nanoscale integration made possible by nanoelectronics.[1]

In the past, research into plasmonics has focused on the noble metals Au and Ag as a result of their high carrier concentrations in the near infrared.[7,8] However, these materials are subject to significant intraband losses at optical frequencies,[9] incompatibility with silicon-based electronics [7] and poor thermal stability,[10] thereby restricting their use in plasmonic devices. Hence the pressing requirement for research into alternative plasmonic materials to overcome these limitations. The plasmonic properties of materials such as TiN,[11] $SrRuO_3$[10] and indium tin oxide (ITO)[12] have previously been investigated; however, while each may have its potential applications, there remains a need for a more comprehensive materials platform in order for the field of plasmonics to be fully integrated into semiconductor devices.

$SrMoO_3$ (SMO) is a compound with a perovskite-type crystal structure which has been shown to exhibit both metallic and Pauli paramagnetic properties.[13] With a bulk cubic crystal cell and lattice constant of 3.975 Å SMO is well-suited for growth on substrates such as $SrTiO_3$ (STO), MgO and $LaAlO_3$ (LAO).[13] Furthermore, the bulk material shows temperature stability up to approximately 1000 K,[14] while thin films of SMO have exhibited low resistivity at room temperature.[13] Such properties are certainly encouraging from the perspective of plasmonics, though reports of the potential use of SMO for plasmonic applications are scarce. In 2000, Mizoguchi et al.[15] reported the measurement of a negative real part of the dielectric permittivity, with a crossover at approximately 720 nm for SMO films grown on glass substrates by reactive DC magnetron sputtering. However, in this instance, the films produced were found to be of relatively poor quality, being both polycrystalline and significantly cracked. More recently, Wadati et al. suggest the existence of plasmon satellites in the HX photoemission spectroscopy data for SMO thin films and therefore recommend further investigations into such properties.[16]

This article explores the potential of SrMoO$_3$ for plasmonic applications. In our initial experiments it was observed that significant variations in the optical properties of SMO samples were present for samples grown on MgO, LAO and STO substrates, under otherwise identical conditions. Such variations were likely a result of changes to the residual strain in the SMO films due to the different lattice parameters of each substrate material. We have explored further this hypothesis by growing SMO films with thicknesses ranging from 40 nm to 100 nm on STO substrates. There is a lattice mismatch of 1.8 % between SMO and STO crystal cells, which results in residual strain in the SMO film. This strain is reduced with the increase of the film thickness. In this paper we report the optical, structural, and electrical properties of SMO thin films as a function of their thicknesses and corresponding residual strain.

## 2. Results

The x-ray diffraction (XRD) pattern acquired for a representative SMO sample is shown in **Figure 1**, in which the narrow (100) and (200) peaks can be considered indicative of a high degree of crystallinity and epitaxial growth. For each sample the out of plane lattice parameter was evaluated from the XRD data, with the resulting values ranging between 3.999 and 4.024 Å, increasing according to the decreasing thickness of the samples.

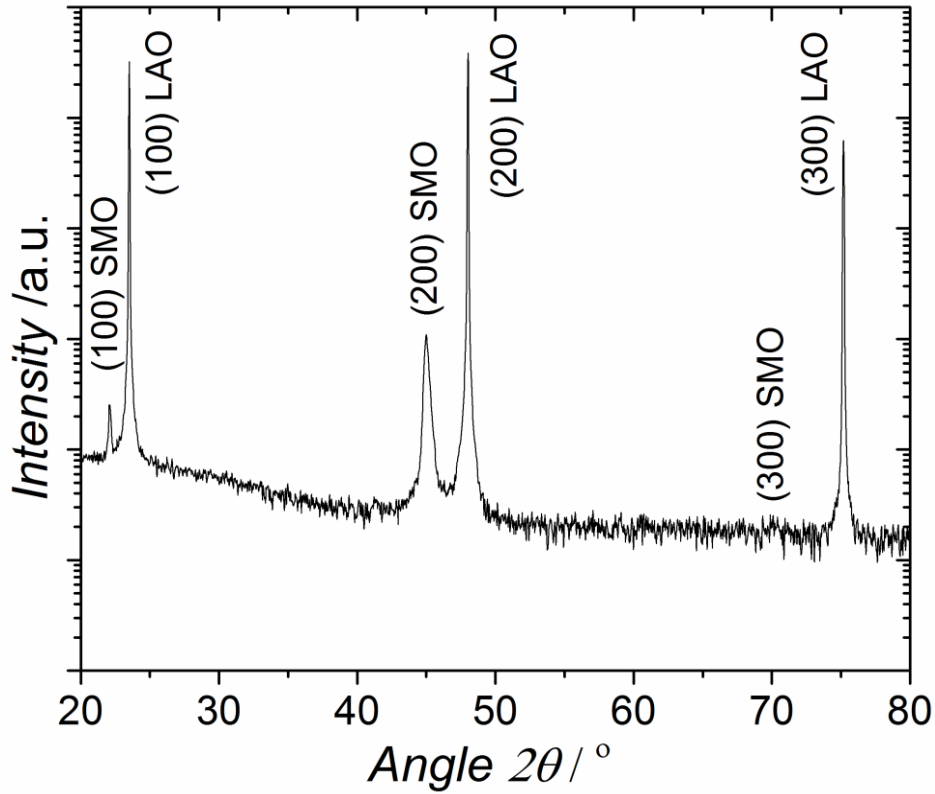

**Figure 1.** Representative XRD pattern of SMO on LAO

The room temperature DC resistivity for each sample was also measured, with values found to range between 100 and 200 μΩ cm according to film thickness as shown in Figure 2. It has been shown previously that the electrical properties of SMO vary according to the choice of substrate as a result of subsequent changes to the lattice parameters,[17] and so it is not surprising that significant variations are observed here. It should be noted however that the resistivity values achieved here, though in close agreement to those of Murakami et al.[18] and Wang et al.,[19] are larger than the 29 μΩ cm reported by Radetinac et al.[20] This is likely attributable to the presence of defects and/or impurities leading to increased scattering rates in the films.[20]

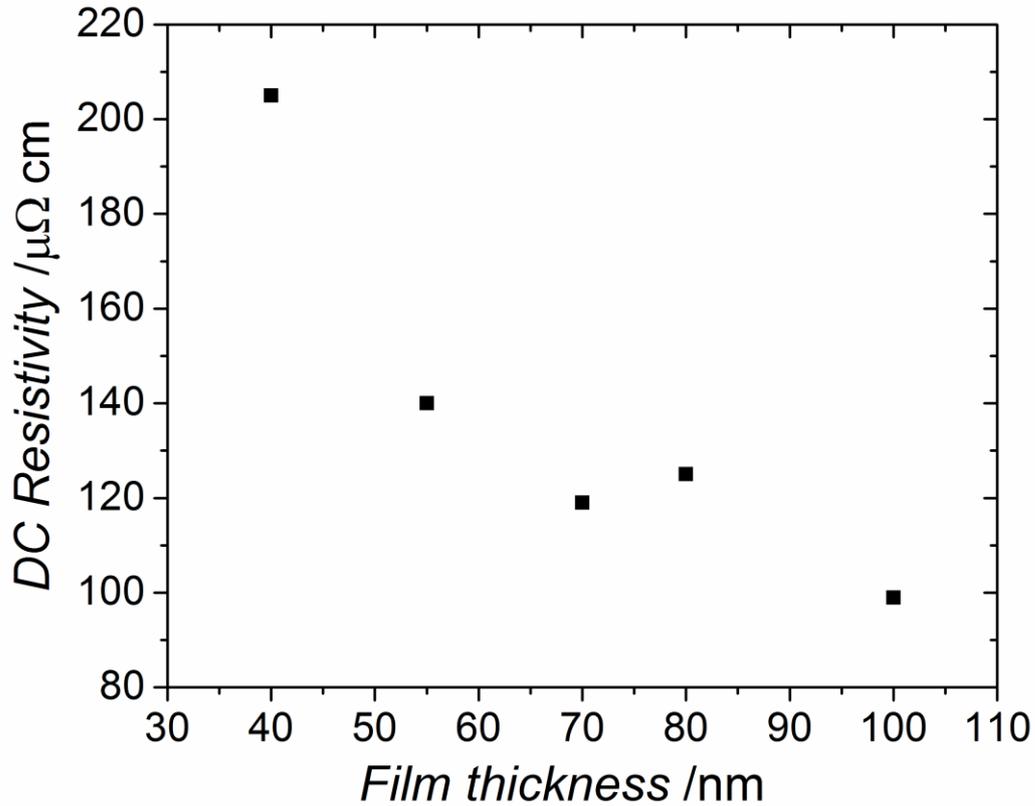

**Figure 2.** Relationship between DC resistivity and SMO film thickness

Figure 3 shows the optical properties for SMO films of various thicknesses grown under the same deposition conditions. The real and imaginary parts of the dielectric permittivity (Figure 3 (a) and (b) respectively) can be seen to follow clear trends according to film thickness. In all cases the results show SMO to be a low loss plasmonic material with crossover from positive to negative real dielectric permittivity (and hence dielectric to plasmonic behaviour) in the visible to near infrared region of the spectrum, agreeing closely with the prior results of Mizoguchi et al.[15] For each sample the optical measurements were in good agreement with the fitted data shown as the mean squared error (MSE) was consistently below 6.5.

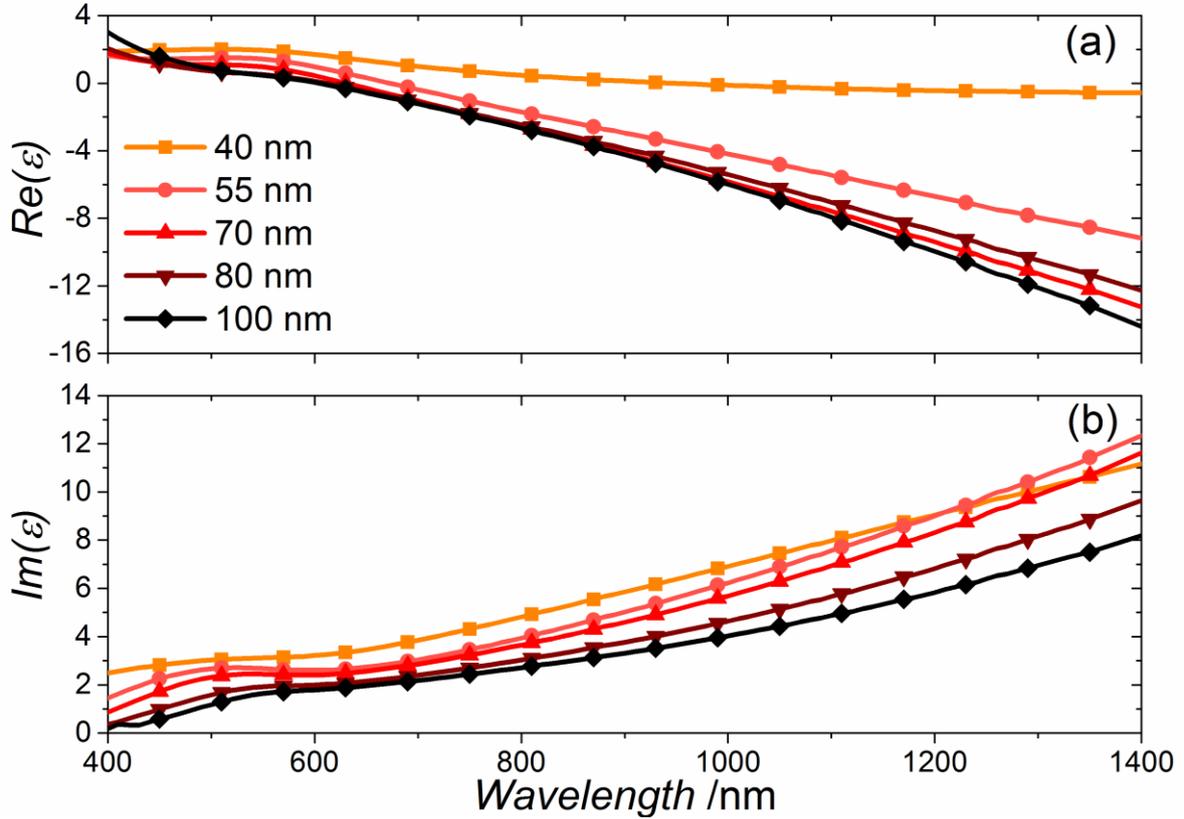

**Figure 3.** Real (a) and Imaginary (b) dielectric permittivity for 40 – 100 nm SMO films on STO substrates

Finally, Hall effect measurements were carried out in order to evaluate the charge carrier concentration and mobility in the samples, found to be of the order of $10^{22}$ cm$^{-3}$ and 1.5 cm$^2$/V s respectively in the case of the 55 nm sample. These results are in good agreement with those of Wang et al.[13]

**3. Discussion**

From the XRD data in Figure 1, together with those of other samples (not shown) we observe that, as the thickness of the samples is increased, the lattice parameter of the SMO tends towards that of the bulk material, ranging from 4.024 to 3.999 Å in the respective cases of the 40 and 100 nm samples. It may therefore be considered that the trends observed in Figure 3 result from changes to the crystal structure of the films. This apparent relationship is shown more clearly in Figure 4, in which the crossover wavelength between positive and negative real dielectric permittivity is plotted against crystal cell lattice parameter, evaluated from the acquired XRD data.

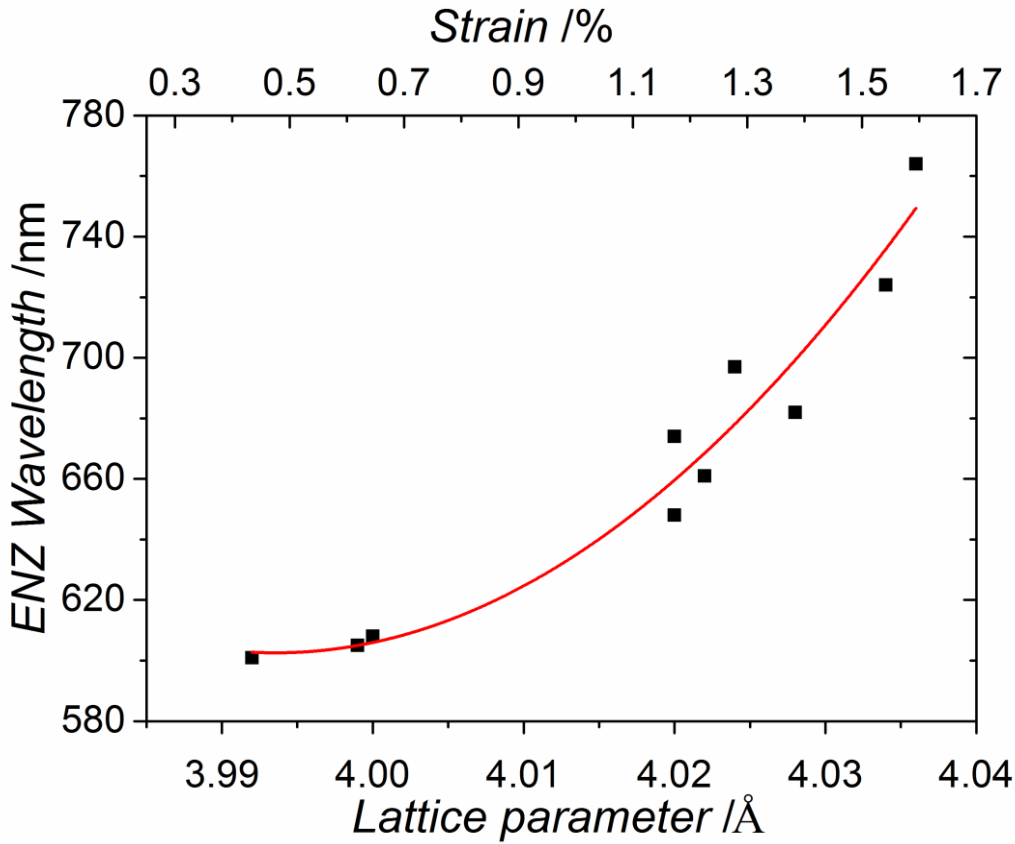

**Figure 4.** Strain dependence of ENZ frequency in SMO films

It can be understood therefore that, as the SMO film thickness is increased, the strain in the film is reduced, as is the DC resistivity (as shown in Figure 2), with both values tending towards those expected of the bulk material. This is reflected in the real part of the dielectric permittivity, as more metallic behaviour can be observed at the film surface as the strain in the film is decreased, with permittivity crossover ranging between 950 and 605 nm, corresponding to 1.31 and 2.05 eV.

From the ellipsometry data and Hall effect measurements it is possible to evaluate the plasma frequency, $\omega_p$, of SMO from the Drude equation.[21]

$$\varepsilon'(\omega) + i\varepsilon''(\omega) = \varepsilon_\infty - \frac{\omega_p^2}{\omega^2 + i\Gamma_D \omega} \qquad (1)$$

Where $\varepsilon_\infty$ is the background dielectric constant and $\Gamma_D$ is the Drude broadening. Equation 1 may be separated into real and imaginary parts to give Equation 2 and 3.[21]

$$\varepsilon'(\omega) = \varepsilon_\infty - \frac{\omega_p^2}{\omega^2 + \Gamma_D^2} \qquad (2)$$

$$\varepsilon''(\omega) = \frac{\Gamma_D \omega_p^2}{\omega^3 + \omega \Gamma_D^2} \qquad (3)$$

Assuming $\varepsilon_\infty$ to be equal to 1, as is the case for a perfectly free-electron-gas,[22] $\omega_p$ can be calculated as approximately 3.8 eV, corresponding to a wavelength of 325 nm, in the case of the 100 nm sample of SMO. Though this is a markedly higher wavelength than the 140 nm (8.9 eV) reported for Au,[22] SMO may still be considered a suitable material for plasmonic applications operating in the visible region of the spectrum.

Finally, the optical properties of SMO were compared with those of existing plasmonic materials. Figure 5 (a) and (b) show respectively the real and imaginary parts of the dielectric permittivity for Au[23], TiN[11] and SMO. It may be observed that, although negative, the real part of the dielectric permittivity (Figure 5 (a)) remains small in magnitude below the crossover frequency. As a result, SMO may be considered well suited to novel transformation optics and metamaterial-based applications. Meanwhile, Figure 5 (b) shows that the optical losses in SMO are significantly lower than those observed in TiN. Furthermore, although the losses are greater than those of gold above approximately 550 nm, the losses at shorter wavelengths are much lower.

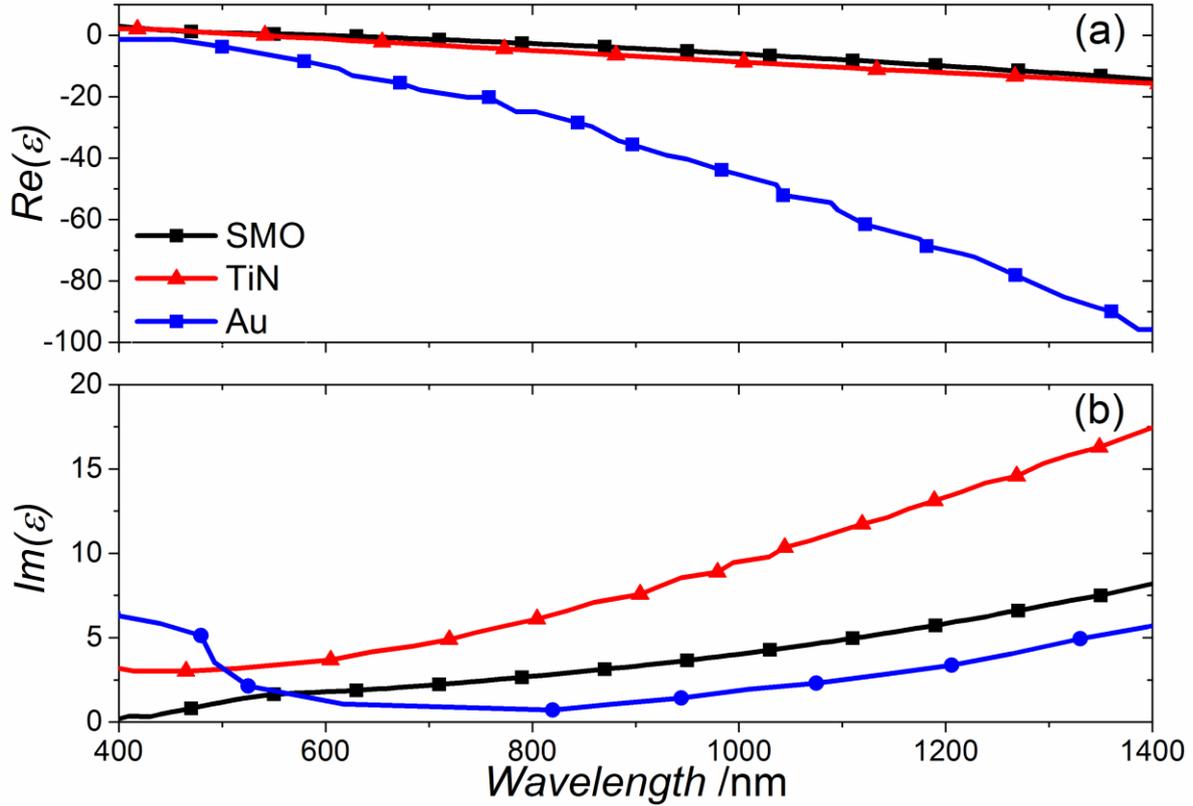

**Figure 5.** Real (a) and Imaginary (b) dielectric permittivity of 100 nm SMO sample, along with reference data for TiN[11] and Au[23]

Table 1 provides a summary of the optical properties of Au, TiN and SMO allowing direct comparison. Here, the wavelengths at which the real part of the dielectric permittivity equals 0 and -2, together with the losses at these frequencies, are shown for each material. Over the wavelengths measured it should be noted that the real part of the dielectric permittivity of Au was always negative. As a result, it is necessary to consider the point at which the permittivity is equal to –2 in order to provide a comparison with the loss characteristics of SMO and TiN. Furthermore, the frequency at which Re($\varepsilon$) = -2 is significant in the consideration of a material for plasmonic applications. According to Mie theory, and assuming variations in Im($\varepsilon$) are small, the polarizability of a particle is subject to a resonant enhancement at approximately this frequency.[24]

**Table 1.** Comparison of wavelengths at which Re(ε) = 0 and Re(ε) = -2 and the losses in each case for Au, TiN and SMO

|     | Wavelength at which Re(ε) = 0 [nm] | Im(ε) when Re(ε) = 0 | Wavelength at which Re(ε) = -2 [nm] | Im(ε) when Re(ε) = -2 |
|-----|-----|-----|-----|-----|
| Au  | -   | -   | 465 | 5.4 |
| TiN | 555 | 3.4 | 652 | 4.2 |
| SMO | 605 | 1.8 | 757 | 2.5 |

Although the losses of Au are lower than those of SMO over a wide frequency range, it can be seen from Table 1 that, at the frequencies of interest, it is Au in which the losses are greater. At the point at which Re(ε) = -2 for each material, the losses present in SMO are approximately 54% lower than those in Au. This is a result of the intraband transition losses present in Au in this frequency range. Similarly, by comparison to TiN, the losses of SMO are approximately 47% lower when Re(ε) = 0 and 40% lower when Re(ε) = -2.

## 4. Conclusion

SrMoO$_3$ films have been grown epitaxially on STO (100) substrates by pulsed laser deposition with thicknesses ranging from 40 to 100 nm. Having studied the films by means of XRD, ellipsometry, and DC resistivity measurements, we have found SMO to exhibit highly tunable optical and electrical properties by means of variations in film thickness resulting in controllable changes to the strain on the film. As a result, we conclude that SrMoO$_3$ represents an extremely promising material for plasmonic and metamaterial applications, owing to its DC resistivity values as low as 100 μΩ cm, together with an ENZ frequency tunable from 1.31 to 2.05 eV and favourable loss characteristics.

## 5. Experimental

The SrMoO$_3$ source material was prepared from SrMoO$_4$ powder supplied by Alfa Aesar (99.9% purity). The powder was ball milled at 300 rpm for 20 hrs in isopropyl alcohol (IPA) before being placed in an oven at 60° C overnight in order to evaporate the IPA. The resulting powder was then reduced in a furnace at 1400° C for 10 hrs under 100 mL min$^{-1}$ gas flow of 5% $H_2$ 95% $N_2$. The powder was pressed into a target and sintered at 1500° C for 12 hours under the same gas flow conditions.

All SMO films were prepared by pulsed laser deposition using a KrF excimer laser (240 nm). In each experiment the laser fluency was 1.2 J cm$^{-2}$, while the laser repetition rate was 8 Hz with a 10 s relaxation period allowed after every 20 pulses. The target material, $SrMoO_3$, was held at a distance of 60 mm from the substrate and was rotated throughout the deposition process. All SMO films were deposited at vacuum conditions, approximately $1\times10^{-7}$ Torr, with the substrate held at a constant temperature of 650° C. After each deposition process the samples were cooled in vacuum to room temperature at a rate of 10° C min$^{-1}$ before removal from the vacuum chamber. Single side polished 5x5 mm STO substrates were used with a thickness of 0.5 mm and (100) orientation. In the case where SMO was deposited on LAO, the substrate was also 0.5 mm thick and (100) orientation.

All characterisations were conducted ex-situ at room temperature. Film thickness measurements were conducted using a Dektak 150 surface profiler. For X-ray diffraction measurements a Bruker D2 PHASER system was used with a Cu $K_\alpha$ wavelength of 1.54 Å. The lattice parameters were subsequently calculated using the Nelson-Riley extrapolation function.[25]

The resistivity of the films was measured using the 4-probe technique. AC Hall effect measurements were achieved using a Lake Shore 8400 Hall system with a field strength of 1.19 T and a current of 15 mA. Optical characterisations were conducted using a J. A. Woollam Co. HS-190 ellipsometer at a 65-80° angle of incidence. The optical constants of the STO and MgO substrates were considered known, while the properties of the $SrMoO_3$, Au, and TiN films were directly fitted to experimental data using the point-by-point method.

**Acknowledgements**

This work was supported by the Engineering and Physical Sciences Research Council (UK). P.K.P. and A.P.M. conceived and designed the research. M.P.W., B.Z. and R.K. carried out the experiments. All authors contributed to the manuscript writing and agreed on its final contents.

Received: ((will be filled in by the editorial staff))

Revised: ((will be filled in by the editorial staff))

Published online: ((will be filled in by the editorial staff))